\documentclass[aps,pra,twocolumn,superscriptaddress,showpacs]{revtex4}
\usepackage{bm}

\usepackage{graphicx}
\usepackage[usenames,dvipsnames]{xcolor}
\usepackage{dcolumn}
\newcommand*{\centt}[1]{\multicolumn{1}{c}{#1}}
\newcommand*{\cent}[1]{\multicolumn{1}{c}{$#1$}}
\newcolumntype{w}[1]{D{.}{.}{#1}}

\begin{document}
\preprint{Version 1.0}

\title{Deuteron and triton magnetic moments from NMR spectra of the hydrogen molecule}

\author{Mariusz Puchalski}
\affiliation{Faculty of Physics, University of Warsaw, Pasteura 5, 02-093 Warsaw, Poland}
\affiliation{Faculty of Chemistry, Adam Mickiewicz University, Umultowska 89b, 61-614 Pozna{\'n}, Poland}

\author{Jacek Komasa}
\affiliation{Faculty of Chemistry, Adam Mickiewicz University, Umultowska 89b, 61-614 Pozna{\'n}, Poland}

\author{Krzysztof Pachucki}
\affiliation{Faculty of Physics, University of Warsaw, Pasteura 5, 02-093 Warsaw, Poland}

\date{\today}

\begin{abstract}
We present a theory and calculations of the nuclear magnetic shielding with finite nuclear mass effects 
and determine magnetic moments of deuteron and triton using the known NMR spectra of HD and HT molecules. 
The results $\mu_d = 0.857\,438\,234\,6(53)\;\mu_N$ and $\mu_t = 2.978\,962\,471(10)\;\mu_N$
are more accurate and in a good agreement with the currently accepted values.
\end{abstract}

\maketitle

When a molecule is placed in a homogeneous magnetic field $\vec B$, its nuclei
experience the field that is shielded by the surrounding electrons $(1-\hat{\sigma})\,\vec B$.
The magnitude of the shielding factor $\sigma$, typically of the order $10^{-5}$,
depends on the relative distance between nuclei, which means that nuclear magnetic resonance 
(NMR) can be used as a tool for obtaining information on the structure of complicated molecules. 
However, when the molecular structure can be calculated,
high-precision NMR spectra can be used to determine the relative magnitude
of nuclear magnetic moments \cite{jaszun_rev}. Namely, the ratio of nuclear magnetic moments
is proportional to the ratio of measured frequencies that flip the nuclear spins
\begin{equation}
\frac{\mu_A\,(1-\sigma_A)}{\mu_B\,(1-\sigma_B)} = \frac{f_A}{f_B}\,\frac{I_A}{I_B}\,. 
\label{01}
\end{equation}
The accuracy to which $\mu_B$ is known can be transferred to $\mu_A$ provided that
one knows with sufficient precision the $\sigma_A-\sigma_B$ difference. Here, we report on 
results of the calculation of this shielding difference for HD and HT  molecules.
Employing these results we determine the magnetic moment of deuteron and triton 
using recent high accuracy measurements of the proton magnetic moment \cite{walz} 
and the ratio of spin flip frequencies \cite{garbacz, neronov1, neronov2, neronov3}. 

Magnetic shielding of the nuclear magnetic moment 
due to the surrounding electrons has been first considered by Ramsey in \cite{ram2} 
with the help of the nonrelativistic Hamiltonian in the external magnetic field. 
His result for the isotropic shielding factor is
\begin{eqnarray}
\sigma^{(0)} &=& \frac{\alpha^2}{3}\,\biggl[
\langle\phi|\sum_{b}\frac{1}{x_b} |\phi\rangle
\label{02}\\&&
+\langle\phi|\sum_a\vec x_a\times\vec p_a
\frac{1}{{\cal E}-H}\,
\sum_{b}\frac{\vec x_b\times\vec p_b}{x_b^3}|\phi\rangle\biggr],
\nonumber
\end{eqnarray}
where $\alpha$ is the fine structure constant, $\vec x_b$ is the position of the electron $b$ 
with respect to the nucleus, $\vec p_b$ is its momentum, $H$ is the molecular Hamiltonian
in the Born-Oppenheimer (BO) approximation, and $\phi$ is the electronic BO wave function. 
An immediate conclusion that can be drawn from this formula is that the shielding of proton and 
deuteron (triton) in the HD (HT) molecule is the same. Upon averaging with the nuclear function $\chi$,
$\sigma^{(0)} = \langle\chi|\sigma^{(0)}(R)|\chi\rangle$, 
the difference in the shielding still vanishes, not only for the ground state but also for 
any excited state. Clearly one has to go beyond the BO approximation and also include 
finite nuclear mass effects in the coupling to the external magnetic field. 
There have been several attempts to calculate the shielding difference in HD (HT), 
and  they are all incorrect or incomplete.  
Neronov and Barzakh in 1977 \cite{neronov1} derived the formula and obtained the result of
\begin{eqnarray}
\delta\sigma({\rm HD}) &\equiv& \sigma_d({\rm HD})-\sigma_p({\rm HD}) = 15.0\cdot 10^{-9} \nonumber \\
\delta\sigma({\rm HT}) &\equiv& \sigma_t({\rm HT})-\sigma_p({\rm HT}) = 20.4\cdot 10^{-9}
\end{eqnarray} 
but they started with the incomplete Hamiltonian, i.e. their formula (4) does not include 
the nuclear spin-orbit interaction (see $g_A-1$ terms in Eq.~(\ref{04}) below). 
Later, the calculations by Jaszu\'nski {\em et al.} \cite{jaszun} simulated nonadiabatic effects
by an artificial charge difference. Their result of $\delta\sigma({\rm HD}) = 9\cdot 10^{-9}$, 
although of the correct magnitude, is not well substantiated from the physical point of view, 
nor is it complete. 
Finally, in the most recent calculations, Golubev and Shchepkin \cite{golubev} 
used more realistic treatment of nonadiabatic effects, 
but their result of $\delta\sigma({\rm HD}) = 9\cdot 10^{-9}$ was also incomplete.
Undoubtedly, the coupling of electron motion to the nuclear motion is partially responsible 
for the difference $\delta\sigma$, but this is not the whole effect.

A derivation of the finite nuclear mass correction to the shielding 
closely follows that of Ref. \cite{ram2, magmol} and starts with the Hamiltonian 
for electrons and nuclei, which includes coupling to the external electromagnetic field
and all possible nucleus $A$ spin-orbit interactions, i.e. ($\hbar=c=1$)
\begin{eqnarray}
H &=& \sum_a\frac{\vec\pi_a^2}{2\,m} + \frac{\vec\pi_A^2}{2\,m_A} + \frac{\vec\pi_B^2}{2\,m_B} + V
-\frac{e_A}{2\,m_A}\,g_A\,\vec I_A\vec B
\nonumber \\ &&
+\sum_b \frac{e_A\,e}{4\,\pi}\,\frac{\vec I_A}{2\,m_A}\cdot
\frac{\vec r_{Ab}}{r_{Ab}^3}\times
\biggl[g_A\,\frac{\vec\pi_b}{m}-(g_A-1)\frac{\vec \pi_A}{m_A}\biggr]
\nonumber \\ &&
+\frac{e_A\,e_B}{4\,\pi}\,\frac{\vec I_A}{2\,m_A}\cdot
\frac{\vec r_{AB}}{r_{AB}^3}\times
\biggl[g_A\,\frac{\vec\pi_B}{m_B}-(g_A-1)\frac{\vec \pi_A}{m_A}\biggr],
\nonumber\\ \label{04}
\end{eqnarray}
where $\vec \pi = \vec p-e\,\vec A$, $\vec A$ is an external magnetic vector potential, 
$g_A$ is a $g$-factor of the nucleus $A$, which is related to the magnetic moment by
$\mu_A = e_A\,g_A\,I_A/(2\,m_A)$, and $V$ is a Coulomb interaction between electrons and nuclei.
In order to derive a formula for the shielding constant, including the finite nuclear mass
corrections, we perform two unitary transformations $\varphi$,
\begin{equation}
\tilde H = e^{-i\,\varphi}\,H\,e^{i\,\varphi}+\partial_t\varphi.
\end{equation}
The first transformation places the gauge origin at the moving nucleus A. 
We assume that the molecule is neutral and that the magnetic field is homogeneous and there is no electric field, so
\begin{eqnarray}
\varphi_1 &=& \sum_a e\,\Bigl(x^i_a\,A^i + \frac{1}{2}\,x^i_a\,x^j_a\,A^i_{,j}\Bigr) 
\nonumber \\ &&
 + e_B\,\Bigl(x^i_B\,A^i + \frac{1}{2}\,x^i_B\,x^j_B\,A^i_{,j}\Bigr) ,
\end{eqnarray}
where $\vec A = \vec A(\vec r_A)$, and $\vec x_a = \vec r_a-\vec r_A$.
The transformed momenta are
\begin{eqnarray}
e^{-i\,\varphi_1}\,\pi^j_a\,e^{i\,\varphi_1} &=& p^j_a+\frac{e_a}{2}\,(\vec x_a\times\vec B)^j\,,\\
e^{-i\,\varphi_1}\,\pi^j_B\,e^{i\,\varphi_1} &=& p^j_B+\frac{e_B}{2}\,(\vec x_B\times\vec B)^j\,,\\
e^{-i\,\varphi_1}\,\pi^j_A\,e^{i\,\varphi_1} &=& p^j_A+\frac{1}{2}\,(\vec D\times\vec B)^j\,,
\end{eqnarray}
where $\vec D = \sum_a e\,\vec x_a + e_B\,\vec x_B$ is the electric dipole moment operator.
We also assume that the total momentum vanishes, thus 
$\vec p_A = -\vec p_B -\sum_a\vec p_a$
and the independent position variables are $\vec x_a$ and $\vec x_B$.
Consider now the electronic Schr\"odinger equation
\begin{equation}
\biggl(\sum_a \frac{\vec p_a^{\,2}}{2\,m} + V - {\cal E}\biggr)\,\phi = 0\,,
\end{equation}
and the next transformation $\varphi_2$ of the form
\begin{equation}
\varphi_2 = -\frac{m}{m_A}\,\sum_a\vec x_a\vec p_B\,,
\end{equation}
which simplifies the nuclear kinetic energy, see Eq. (\ref{14}) below.
The electron momenta are changed to
\begin{equation}
p_a^{\prime i} = e^{-i\,\varphi_2}\,p_a^i\,e^{i\,\varphi_2} = p_a^i -\frac{m}{m_A}\,p_B^i\,, 
\end{equation}
and the potential
\begin{eqnarray}
V' -{\cal E} &=& 
e^{-i\,\varphi_2}\,(V - {\cal E})\,e^{i\,\varphi_2} 
\nonumber \\ &\approx& 
V-{\cal E}+\frac{m}{m_A}\,\sum_a\vec x_a\vec \nabla_B(V-{\cal E})\,,
\end{eqnarray}
where we omitted higher order terms in the electron nucleus mass ratio.
The new Hamiltonian $\tilde H$ after both transformations 
with $\vec p_{\rm el} = \sum_a \vec p_a$ and $\vec x_{\rm el} = \sum_a \vec x_a$  becomes
\begin{widetext}
\begin{eqnarray}
\tilde H &=& \sum_a\frac{1}{2\,m}\,
\Bigl(\vec p_a+\frac{e}{2}\,\vec x_a\times\vec B-\frac{m}{m_A}\,\vec p_B\Bigr)^2
+\frac{1}{2\,m_B}\,\Bigl(\vec p_B+\frac{e_B}{2}\,\vec x_B\times\vec B\Bigr)^2
\nonumber \\ &&
+\frac{1}{2\,m_A}\,\biggl(\vec p_B+\vec p_{\rm el}-\frac{\vec D\times\vec B}{2}\biggr)^2
+V+\frac{m}{m_A}\,\sum_a\vec x_a\vec \nabla_B (V - {\cal E})
-\frac{e_A}{2\,m_A}\,g_A\,\vec I_A\,\vec B
\nonumber \\ &&
-\sum_a\frac{e_A\,e}{4\,\pi}\,\frac{\vec I_A}{2\,m_A}\cdot
\frac{\vec x_a}{x_a^3}\times\biggl[\frac{g_A}{m}\,\biggl(
\vec p_a+\frac{e}{2}\,\vec x_a\times\vec B-\frac{m}{m_A}\,\vec p_B\biggr)
+\frac{(g_A-1)}{m_A}\,\biggl(\vec p_B+\vec p_{\rm el}-\frac{\vec D\times\vec B}{2}\biggr)\biggr]
\nonumber \\ &&
-\frac{e_A\,e_B}{4\,\pi}\,\frac{\vec I_A}{2\,m_A}\cdot
\frac{\vec x_B}{x_B^3}\times\biggl[\frac{g_A}{m_B}\,\biggl(
\vec p_B+\frac{e_B}{2}\,\vec x_B\times\vec B\biggr)
+\frac{(g_A-1)}{m_A}\,\biggl(\vec p_B+\vec p_{\rm el}-\frac{\vec D\times\vec B}{2}\biggr)\biggr]\,.
\label{13}
\end{eqnarray}
\end{widetext}
From this Hamiltonian one derives the shielding defined by
\begin{equation}
H_{\rm eff} = -\frac{e_A}{2\,m_A}\,g_A\,\vec I_A\,(1-\hat\sigma(\vec R))\vec B\,.
\label{91}
\end{equation}
Let us consider firstly the nonrelativistic Hamiltonian
$H_{\rm nrel} = H + H_{\rm n}$, where
\begin{equation}
H = \sum_a \frac{\vec p_a^{\;2}}{2\,m} + V, 
\end{equation}
with
\begin{equation}
V = \frac{1}{R} + \frac{1}{x_{12}} - \frac{1}{x_1} - \frac{1}{x_2} - \frac{1}{|\vec x_1+\vec R|} - \frac{1}{|\vec x_2+\vec R|}, 
\end{equation}
and where
\begin{equation}
H_{\rm n} =
\frac{\vec p_B^{\;2}}{2\,m_{\rm n}}
+\frac{\bigl(\sum_a\vec p_a\bigr)^2}{2\,m_A} 
+ \frac{m}{m_A}\, \sum_a\vec x_a\cdot\vec\nabla_B(V-{\cal E}),
\nonumber \\ \label{14}
\end{equation}
with $m_{\rm n} = m_A\,m_B/(m_A+m_B)$ being the nuclear reduced mass.
We assume the Born-Oppenheimer approximation and include $H_{\rm n}$ perturbatively 
using nonadiabatic perturbation theory (NAPT) \cite{napt}. 
In the zeroth order, the shielding is given by Eq. (\ref{02}).
The leading nonadiabatic correction to the shielding $\hat\sigma^{(1)}$ 
is linear in the electron-nuclear mass ratio, and we split it into four parts \cite{magmol}:
$\sigma^{(1)} = \sigma_{\rm n} + \sigma_{\rm d} + \sigma_{\rm s} + \sigma_{\rm l}$.
The explicit formulae for the isotropic shielding in 
H$_2$ and isotopomers: $e_A=e_B=-e$, with
$\vec R = \vec r_{AB} = \vec r_A - \vec r_B = -\vec x_B$ are
\begin{widetext}
\begin{eqnarray}
\sigma_{\rm n} &=&\frac{\alpha^2}{3}\,\biggl[
2\,\langle\phi|\sum_{b}\frac{1}{x_b}\,\frac{1}{({\cal E}-H)'}
\,\stackrel{\leftrightarrow}{H_{\rm n}} |\phi\rangle
+\langle\phi|\sum_a\vec x_a\times\vec p_a
\frac{1}{({\cal E}-H)'}
\,(\stackrel{\leftrightarrow}{H_{\rm n}}-{\cal E}_{\rm a})
\,\frac{1}{({\cal E}-H)'}\,
\sum_{b}\frac{\vec x_b\times\vec p_b}{x_b^3}|\phi\rangle
\nonumber \\ &&
+\langle\phi|
\stackrel{\leftrightarrow}{H_{\rm n}}
\,\frac{1}{({\cal E}-H)'}\,
\sum_a\vec x_a\times\vec p_a
\frac{1}{({\cal E}-H)'}
\sum_{b}\frac{\vec x_b\times\vec p_b}{x_b^3}|\phi\rangle
\nonumber \\ &&
+\langle\phi|\sum_a\vec x_a\times\vec p_a
\,\frac{1}{({\cal E}-H)'}\,
\sum_{b}\frac{\vec x_b\times\vec p_b}{x_b^3}\,
\frac{1}{({\cal E}-H)'}
\,\stackrel{\leftrightarrow}{H_{\rm n}}|\phi\rangle\biggr], \label{19}\\
\sigma_{\rm d} &=& \frac{\alpha^2}{3}\,\langle\phi|
\frac{1}{m_B\,R}-\frac{1}{m_A}\,\,\frac{(g_A-1)}{g_A}\,
\biggl(\sum_b\frac{\vec x_b}{x_b^3}+\frac{\vec R}{R^3}\biggr)\,(\vec x_{\rm el}+\vec R)
|\phi\rangle\,, \label{20}\\
\sigma_{\rm s} &=& 
\frac{\alpha^2}{3}\,\langle\phi|\sum_a\vec x_a\times\vec p_a\,
\frac{1}{({\cal E}-H)'}
\biggl\{
-\frac{1}{m_{\rm n}} \, \frac{\vec R\times\vec P}{R^3}
+\frac{1}{m_A}\,\biggl(\sum_{b}\frac{\vec x_b}{x_b^3}+\frac{\vec R}{R^3}\biggr)\times
\biggl[\vec P-\frac{(g_A-1)}{g_A}\,(\vec P - \vec p_{\rm el}) \biggr]
\biggr\}|\phi\rangle\,,\label{21}\\
\sigma_{\rm l} &=&\frac{\alpha^2}{3} \langle\phi|
\sum_b\frac{\vec x_b\times\vec p_b}{x_b^3}
\frac{1}{({\cal E}-H)'}
\biggl[ 
\frac{1}{m_A}\,\bigl(\vec x_{\rm el}+\vec R\bigr)\times
\bigl(2\,\vec P-\vec p_{\rm el}\bigr)
-\frac{1}{m_{\rm n}}\,\vec R\times\vec P\biggr]|\phi\rangle\,, \label{22} 
\end{eqnarray}
\end{widetext}
where $\vec P = -i\,\vec\nabla_R$, and
$\langle\phi|\stackrel{\leftrightarrow}{\Delta}_R
|\psi\rangle = 
-\langle\vec\nabla_R\phi|
\vec\nabla_R\psi\rangle$.
The result in Ref. \cite{magmol} contains a mistake in $\sigma_{\rm d}$, which we correct here.

In the numerical evaluation of Eqs.~(\ref{19}-\ref{22}),
the ground state wave function is represented with the explicitly correlated Gaussian functions of the form
\begin{equation}
  \phi_{\Sigma^+} = e^{-a_{1A}\,r_{1A}^2 -a_{1B}\,r_{1B}^2 -a_{2A}\,r_{2A}^2 -a_{2B}\,r_{2B}^2 - a_{12}\,r_{12}^2 }. \\ 
\end{equation}
The resolvent $1/({\cal E}-H)$ includes the sum of 
$\Sigma^-$ and $\Pi$ states, which are represented as
\begin{eqnarray} 
  \phi_{\Sigma^-} &=& \vec R \cdot (\vec r_{1A} \times \vec r_{2A}) \, \phi_{\Sigma^+} \\ 
 \vec\phi_{\Pi} &=& \vec R \times \vec r_{1A}\,\phi_{\Sigma^+}.
\end{eqnarray}
All of the above matrix elements are in electronic variables,
so the derivatives with respect to the internuclear distance
have to be obtained in advance. For this purpose we use the fact
that total angular momentum vanishes on $\phi$, so
\begin{equation}
   \vec R \times \vec \nabla_R | \phi \rangle = 
   - \Big(\sum_b \vec x_b \times \vec \nabla_b \Big)  | \phi \rangle, 
   \end{equation}
and from the $R$-derivative of the Schr\"odinger equation in the BO approximation we get
\begin{equation}
\vec \nabla_R  |\phi\rangle = \frac{1}{({\cal E}-H)'} \,\vec \nabla_R(V)|\phi\rangle\,.
\end{equation}
Since we calculate only the shielding difference, 
all of the terms with the reduced mass, including the most difficult $\Delta_R$, are omitted.
Calculations are performed using 128 and 256 basis function
for each symmetry with global optimization of all nonlinear parameters.
Numerical results in the range $R\in\langle 0,6\rangle$ a.u. are presented in Table~\ref{Tres}.
\squeezetable
\begin{table}[!htb]
 \caption{\label{Tres}
Shielding as a function of the internuclear distance $R$. 
Results to be multiplied by $10^{-6}$. The numerical uncertainty of $\sigma^{(0)}$ is $(20)$, 
while of $\delta\sigma$ is $(2)$ on last digits. }
 \begin{ruledtabular}
 \begin{tabular}{l@{\extracolsep{\fill}}*{ 2}{w{3.11}}@{\extracolsep{\fill}}*{ 2}{w{3.11}}}
$R$[a.u.]& \centt{$\sigma^{(0)}$}  & \centt{$\delta\sigma({\rm HD})$} & \centt{$\delta\sigma({\rm HT})$}  \\
 \hline
0.00 & 59.936\,77  &  \cent{\infty} &  \cent{\infty} \\
0.10 & 58.308\,74  &  0.053\,997 &  0.069\,030 \\
0.20 & 54.933\,43  &  0.030\,630 &  0.038\,311 \\
0.40 & 47.435\,08  &  0.020\,577 &  0.025\,604 \\
0.60 & 41.035\,38  &  0.018\,685 &  0.023\,329 \\
0.80 & 36.014\,57  &  0.018\,680 &  0.023\,243 \\
1.00 & 32.118\,10  &  0.019\,197 &  0.023\,663 \\
1.10 & 30.504\,37  &  0.019\,486 &  0.023\,874 \\
1.20 & 29.074\,46  &  0.019\,759 &  0.024\,044 \\
1.30 & 27.804\,09  &  0.019\,991 &  0.024\,153 \\
1.40 & 26.672\,56  &  0.020\,181 &  0.024\,201 \\
1.50 & 25.662\,62  &  0.020\,324 & 0.024\,186 \\
1.60 & 24.759\,25  &  0.020\,416 & 0.024\,107 \\
1.70 & 23.949\,94  &  0.020\,464 & 0.023\,973 \\
1.80 & 23.224\,07  &  0.020\,469 & 0.023\,792 \\
1.90 & 22.572\,23  &  0.020\,436 & 0.023\,569 \\
2.00 & 21.986\,84  &  0.020\,367 & 0.023\,308 \\
2.10 & 21.461\,05  &  0.020\,261 & 0.023\,015 \\
2.20 & 20.988\,70  &  0.020\,131 & 0.022\,700 \\
2.30 & 20.565\,01  &  0.019\,973 & 0.022\,362 \\
2.40 & 20.185\,30  &  0.019\,792 & 0.022\,010 \\
2.50 & 19.845\,67  &  0.019\,587 & 0.021\,643 \\
2.60 & 19.542\,83  &  0.019\,361 & 0.021\,265 \\
2.70 & 19.273\,36  &  0.019\,117 & 0.020\,880 \\
2.80 & 19.034\,52  &  0.018\,855 & 0.020\,488 \\
2.90 & 18.823\,95  &  0.018\,570 & 0.020\,085 \\
3.00 & 18.638\,83  &  0.018\,280 & 0.019\,688 \\
3.20 & 18.337\,14  &  0.017\,652 & 0.018\,886 \\
3.40 & 18.113\,34  &  0.016\,989 & 0.018\,092 \\
3.60 & 17.952\,68  &  0.016\,308 & 0.017\,322 \\
3.80 & 17.842\,09  &  0.015\,640 & 0.016\,598 \\
4.00 & 17.769\,75  &  0.015\,002 & 0.015\,930 \\
4.20 & 17.725\,68  &  0.014\,405 & 0.015\,320 \\
4.40 & 17.701\,48  &  0.013\,866 & 0.014\,780 \\
4.60 & 17.690\,67  &  0.013\,394 & 0.014\,311 \\
4.80 & 17.688\,47  &  0.012\,976 & 0.013\,899 \\
5.00 & 17.691\,28  &  0.012\,632 & 0.013\,557 \\
5.20 & 17.696\,74  &  0.012\,344 & 0.013\,270 \\
5.40 & 17.703\,34  &  0.012\,099 & 0.013\,023 \\
5.60 & 17.710\,10  &  0.011\,889 & 0.012\,810 \\
5.80 & 17.716\,48  &  0.011\,724 & 0.012\,638 \\
6.00 & 17.722\,24  &  0.011\,584 & 0.012\,489 \\
$\infty$ 
     & 17.750\,45  &   0.010\,753 & 0.011\,326
 \end{tabular}
 \end{ruledtabular}
\end{table}
Although the results for $R>3$ are not in principle needed,
we use them for testing against the known separated atoms limit, which is
\begin{equation}
\sigma^{(1)}(R=\infty) = -\frac{\alpha^2}{3}\,\frac{m}{M}\,
\biggl(1+\frac{g_A-1}{g_A}\biggr)\,.
\end{equation}
On the other hand, at small $R$ we observe $1/R$ behavior, 
which is not an artifact but a result of the shielding of the nucleus $A$ by the nucleus $B$,
see Fig. 1.
\begin{figure}
\caption{The difference $\delta\sigma({\rm HD},R)$ in ppm 
         of the shielding constant between the deuteron and the proton in HD 
         as a function of the internuclear distance $R$. 
         The horizontal line is a separated atoms limit, while the dotted line
         is the $1/R$ asymptotics that comes from the direct interaction between nuclei.}
\centerline{\includegraphics[scale=0.48]{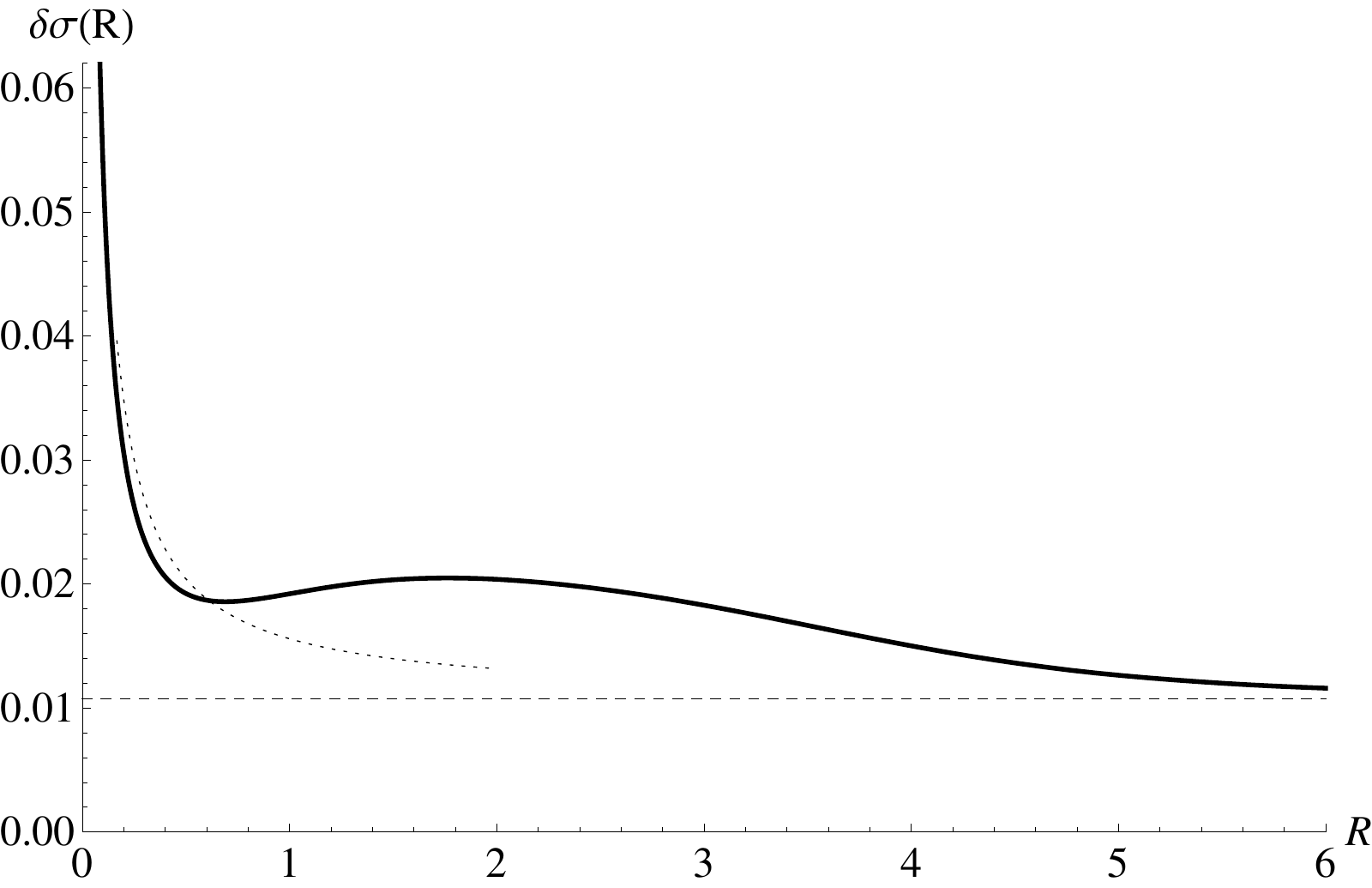}}
\end{figure}
Our results at $R=1.4$ a.u. in comparison to the known previous calculations
are presented in Table \ref{Tcomp}
\begin{table}[!htb]
 \caption{\label{Tcomp}
Extrapolation to a complete basis and comparison with the previous calculations 
of the isotropic nonrelativistic BO shielding
in HD (HT) at $R=1.4$ a.u. Results to be multiplied by $10^{-6}$.}
 \begin{ruledtabular}
 \begin{tabular}{r w{3.11} w{3.11} w{3.11}}
size/ref.& \cent{\sigma^{(0)}_{\rm el}}  & \cent{\delta \sigma({\rm HD})} & \cent{\delta \sigma({\rm HT})} \\ \hline
128      & 26.673\,193                        &  0.020\,188\,2    & 0.024\,214\,5  \\ 
256      & 26.672\,556                        &  0.020\,181\,2  & 0.024\,201\,0   \\
512      & 26.672\,422                        &  0.020\,179\,9    & 0.024\,199\,6  \\
$\infty$ & 26.672\,387(35)                    &  0.020\,179\,6(3) & 0.024\,199\,4(2)  \\[1ex]
\cite{komasa} (1995) & 26.813\,9  \\
\cite{gauss1} (1996) & 26.680     \\
\cite{jaszun} (2011) & 26.677\,111 \\
 \end{tabular}
 \end{ruledtabular}
\end{table}

The total isotropic magnetic shielding $\sigma$
is obtained by averaging with the nuclear wave function $\chi$,
$\sigma = \langle\chi|\sigma^{(0)}(R) +\sigma^{(1)}(R)|\chi\rangle$,
where $\chi$ is a solution of the nuclear radial equation with the BO potential 
augmented by adiabatic correction.
Since the measurement \cite{garbacz} was performed at $T=300$~K we include contributions from
the excited rotational states up to $J=9$ according to Boltzmann distribution.
The averaged result for the (Ramsey) shielding factor of the H$_2$ and isotopomers is
$\sigma^{(0)}  = 26.335\,17(20)\cdot 10^{-6}$.
However, $\sigma^{(0)}$ does not include relativistic corrections that are 
of the relative order of $\alpha^2 \sim 10^{-4}$. The related estimate 
from Ref. \cite{jaszun} is much smaller
than the relativistic correction to the shielding of the individual hydrogen atoms,
so it is possibly incorrect.
\begin{table}[!htb]
 \caption{\label{Tresults} Determination of the deuteron and the triton magnetic moments. $\mu_p({\rm HD})$ 
                           denotes the shielded magnetic moment of the proton in HD molecule.}
 \begin{ruledtabular}
 \begin{tabular}{l w{1.19} r}
                                  & \centt{value}            & \centt{Ref.}  \\ \hline
$\mu_p$                            & 2.792\,847\,350(9)\;\mu_N &\cite{walz} \\
$\delta\sigma({\rm HD},T=300K)$     &  0.020\,20(2)\cdot 10^{-6} & this work \\
$\mu_p({\rm HD})/\mu_d({\rm HD})$ & 3.257\,199\,514(21) & \cite{garbacz} \\
                                    & 3.257\,199\,531(29)&\cite{neronov2} \\
                                    & 3.257\,199\,520(17)& averaged       \\
$\mu_d = \mu_d({\rm HD})/\mu_p({\rm HD})\,(1+\delta\sigma)\,\mu_p$ & 0.857\,438\,234\,6(53)\,\mu_N & this work\\
                                              & 0.857\,438\,230\,8(72)\,\mu_N & \cite{codata}\\[2ex]
$\delta\sigma({\rm HT},T=300K)$ &  0.024\,14(2)\cdot 10^{-6} & this work \\
$\mu_t({\rm HT})/\mu_p({\rm HT})$ & 1.066\,639\,893\,3(7)     &\cite{neronov3}\\
$\mu_t = \mu_t({\rm HT})/\mu_p({\rm HT})\,(1+\delta\sigma)$ &2.978\,962\,471(10)\,\mu_N & this work \\
                                       &2.978\,962\,448(38)\,\mu_N & \cite{codata} 
 \end{tabular}
 \end{ruledtabular}
\end{table}
The shielding differences in HD and HT are presented in Table \ref{Tresults}.
Their uncertainties come from the unknown higher order nonadiabatic corrections.
From these differences and from Eq. (\ref{01}) we obtain the magnetic moments
of deuteron and triton, see Table \ref{Tresults}.
The ratio of shielded magnetic moments for HD is the average of two independent 
and consistent measurements \cite{neronov1,neronov2} and \cite{garbacz}.
The obtained result for $\mu_d$ is more accurate and in good agreement with the presently accepted value, 
which was obtained from an unpublished experimental 
result by Philips {\em et al.} (1984), see \cite{codata} for details.
Regarding $\mu_t$, the CODATA \cite{codata} value is based on the earlier, 
less accurate work \cite{neronov2}, while we use a more recent one \cite{neronov3} 
and correct the value for $\delta\sigma$, which leads to even smaller uncertainty of $\mu_t$. 
We should note however, that the result of \cite{neronov3} needs confirmation 
and because of the lack of information on temperature, we assumed $T=300$~K. 

In summary, we have determined improved values for the deuteron and triton magnetic moments.
This demonstrates that NMR spectroscopy combined with precise calculations of the shielding factor
may lead to the most accurate determination of nuclear magnetic moments.
Moreover, one may consider improving the determination of magnetic moments of other light nuclei, 
such as $^3$He, since the shielding factor for molecular hydrogen can be calculated including 
nonadiabatic and relativistic effects as has been already done for $^3$He atom \cite{rudzins}.

\section*{Acknowledgments}
We would like to thank Piotr Garbacz for bringing our attention to his excellent experimental results, and
we gratefully acknowledge interesting discussions with
Micha\l\ Jaszu\'nski and Grzegorz \L ach. This work was supported
by Polish National Center for Science grant no. 2012/04/A/ST2/00105,
and by a computing grant from Pozna\'n Supercomputing and Networking Center and by PL-Grid Infrastructure.

\end{document}